# Highly Confined Optical Modes in Nanoscale Metal-Dielectric Multilayers

Ivan Avrutsky[1*], Ildar Salakhutdinov[1], Justin Elser[2], Viktor Podolskiy[2]
[1]*Department of Electrical and Computer Engineering, Wayne State University, Detroit, Michigan 48202*
[2]*Department of Physics, Oregon State University, Corvallis, Oregon 97331*

We show that a stack of metal-dielectric nanolayers, in addition to the long- and short-range plasmon polaritons, guides also an entire family of modes strongly confined within the multilayer – the bulk plasmon polariton modes. We propose a classification scheme that reflects specific properties of these modes. We report experimental verification of the bulk modes by measuring modal indices in a structure made of three pairs of silica(~29nm)/gold(~25nm) layers.

Nanoscale confinement of light is of great interest for applications in sensing, imaging, all-optical signal processing and computing. Subwavelength confinement attributed to gap plasmon polaritons (GPP) has been demonstrated in a thin dielectric layer surrounded by metallic claddings.[1] Here we present another solution to subdiffraction confinement of light and show that a stack of metal-dielectric nanolayers guides a family of modes strongly confined within the multilayer – the bulk plasmon polariton (BPP) modes. The bulk modes have very short penetration depth into the claddings even if the claddings are made of dielectric materials. Their modal indices (ratio of the light velocity in vacuum to the phase velocity of a guided mode) are typically large in absolute value and may be both positive and negative.[2] We propose a classification scheme that reflects specific properties of BPPs. We verify BPPs experimentally by measuring their modal indices in a structure made of three pairs of silica/gold nano-layers.

When considering the light confinement in a waveguide, the modal index $n^*$ (rather than group index) is of interest because it defines the modal profile. The field penetration length into the cladding with permittivity $\varepsilon_{cl}$ is $L_{cl}=\lambda/2\pi\sqrt{n^{*2}-\varepsilon_{cl}}$, where $\lambda$ is the vacuum wavelength. For given $\varepsilon_{cl}$, the larger modal index leads to shorter penetration length and stronger confinement. This justifies interest to the high-index modes. In all-dielectric waveguides, the modal index is smaller than the core index, which limits the confinement scale to ~$\lambda$/7 in a silicon-on-insulator waveguide with silicon core ($n\approx3.5$) and silica or vacuum claddings.[3]

Surface plasmon polariton (SPP)[4] propagating along the interface between media with permittivities $\varepsilon_m$ and $\varepsilon_d$ of different sign (metal and dielectric) is an example of a strongly confined mode. Its modal index $n^*_{SPP}=\sqrt{\varepsilon_d\varepsilon_m/(\varepsilon_d+\varepsilon_m)}$ is slightly above the index of the dielectric. In the visible and near infrared spectral region, the permittivity of metal is typically negative and $|\varepsilon_m|>>\varepsilon_d$ leading to subwavelength field penetration into metal, while the penetration into the dielectric can be as large as several wavelengths. A remarkable exception is the case of resonant SPPs[5] when permittivities of materials at different sides of the interface are exactly opposite: $\varepsilon_m+\varepsilon_d\to0$ and thus $n^*\to\infty$. In homogeneous media, the field distribution of the resonant SPP is expected to be confined infinitely close to the interface. Actual scale of the field distribution is defined by the applicability of the concept of dielectric permittivity, that is, by the discrete atomic structure of materials. At optical frequencies, the resonant SPPs are possible if the dielectric has huge optical gain.[6]

A metallic film of thickness $t_m$ between dielectric claddings supports SPPs at each metal-dielectric interface. In sufficiently thin films, coupling between individual SPPs leads to formation of symmetric and anti-symmetric film modes known as long- and short-range plasmon polaritons (LRP and SRP).[7, 8] The coupling leads to splitting of the dispersion characteristics of these modes. The modal index of the LRP $n^*_{LRP}\approx\sqrt{\varepsilon_d+(\pi t_m/\lambda)^2((\varepsilon_d-\varepsilon_m)\varepsilon_d/\varepsilon_m)^2}$ is reduced compared to that of SPP approaching the cladding index $n_d\approx\sqrt{\varepsilon_d}$. Its penetration into the claddings increases accordingly. The modal index of the SRP is higher than that of SPP: $n^*_{SRP}\approx\sqrt{\varepsilon_d+(\lambda/\pi t_m)^2(\varepsilon_d/\varepsilon_m)^2}$, yielding the confinement scale $L_{SRP}=t_m+2L_{cl}\approx t_m(1+|\varepsilon_m|/\varepsilon_d)$. Note that even for nanoscale (~10nm) metal films, both LRP and SRP are typically extended to quite a macroscopic scale. The modal indices of SRP and LRP in fused silica($\varepsilon_d$=1.444$^2\approx$2.085)[9]/gold($\varepsilon_m$=-114.5+ 11.01$i$)[10] system at $\lambda$=1550nm as functions of metal thickness are shown in the first column of Fig. 1.

The splitting of modal indices resulting from the coupling is an electromagnetic analog of a universal phenomenon appearing in linear wave-dynamic systems: energy level splitting in quantum mechanics, characteristic frequencies splitting in acoustics, splitting of relaxation time in coupled ac circuits, etc. Modes in a metal-dielectric multilayer can be represented as linear combination of surface modes at individual interfaces, with splitting between modal indices determined by the overlapping of original wavefunctions.

The system of two metal films supports four modes, which differ by symmetry. Two of these modes are dominated by SPPs propagating at interfaces with the claddings, and their properties are identical to SRP and LRP. The other two modes are primarily composed of SPPs propagating inside the gap between the metal films. The mode with anti-symmetric magnetic field distribution experiences the cut-off when the gap thickness becomes less than $d_c\approx\lambda/2\sqrt{\varepsilon_d}$, and therefore it does not exist in nanoscale structures. The mode with symmetric field distribution, the GPP, survives. For a nanoscale dielectric gap of thickness $t_d$ between infinitely thick metallic claddings, the modal index $n^*_{GPP}$ of gap plasmon polariton

$$n^*_{GPP} \approx \sqrt{\varepsilon_d + \frac{1}{2}\left(\frac{\lambda}{\pi t_d}\frac{\varepsilon_d}{\varepsilon_m}\right)^2 + \sqrt{\left(\frac{\lambda}{\pi t_d}\frac{\varepsilon_d}{\varepsilon_m}\right)^2 (\varepsilon_d - \varepsilon_m) + \frac{1}{4}\left(\frac{\lambda}{\pi t_d}\frac{\varepsilon_d}{\varepsilon_m}\right)^4}} \quad (1)$$

is inversely proportional to the gap size. Note that the modal index of the gap plasmon polariton greatly exceeds that of the SRP, providing truly nanoscale mode confinement. Similar modes exist also in a gap between sharp metallic wedges.[11] Exact dispersion equations for modes in a gap between metallic claddings have been proposed earlier.[12]

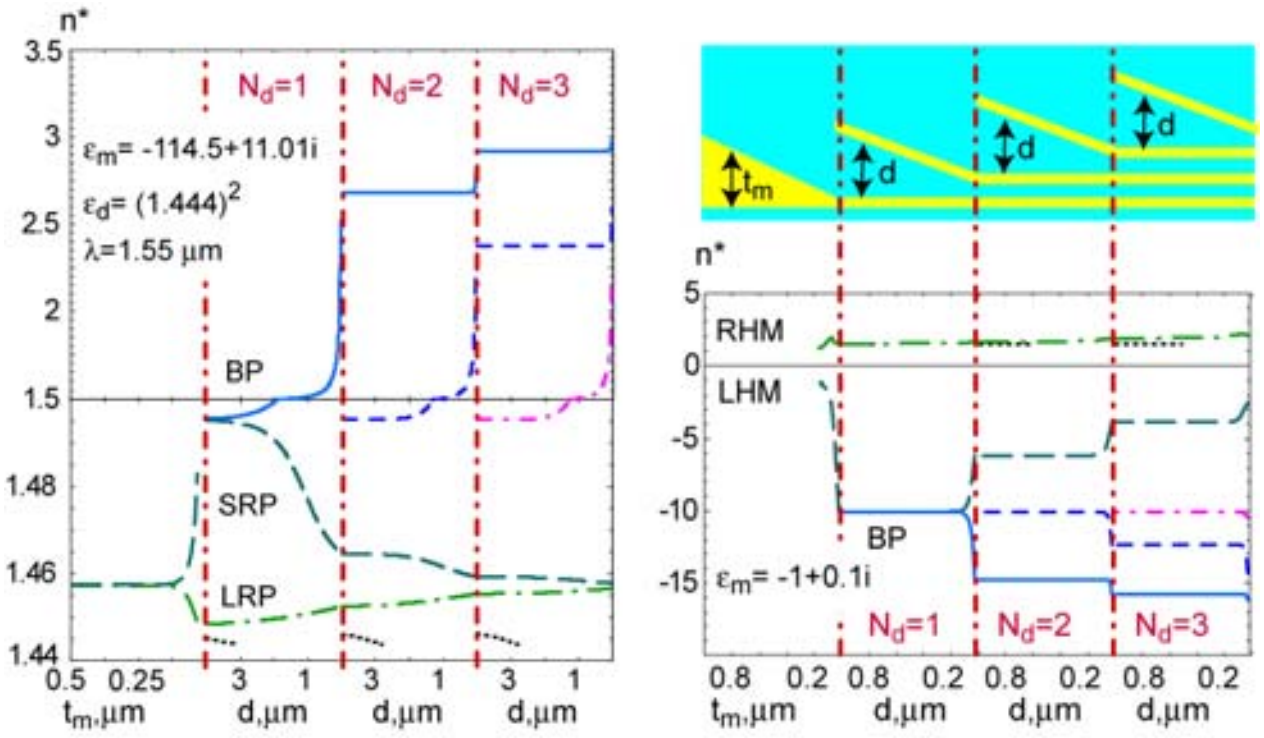

FIG. 1. (Color online) Evolution of coupled SPP modal indices in the layered materials with dielectric claddings for $|\varepsilon_m|>\varepsilon_d$ (left) and $|\varepsilon_m|<\varepsilon_d$ (bottom right). System geometries are sketched in top-right inset, where blue (gray) and yellow (light-gray) regions represent dielectric and metal layers. Graphs at the left correspond to Au/$SiO_2$ composite considered in this work; graphs at the right represent its "negative index" counterpart with $\varepsilon_m = -1+0.1i$ and $\varepsilon_d=1.444^2$; λ=1550nm. First vertical portion of the figure illustrates the appearance of SRP/LRP doublet as a result of splitting between two SPPs (left) and as a result of mode/anti-mode formation (right). Other vertical portions illustrate modal indices in a layered structure when a metal layer is being gradually brought closer to the nanolayer stack.

As more layers are added to the structure, the collective interaction of SPPs leads to the mutual repulsion of modal indices. The gap plasmon polaritons contribute to the emergence of high-index supermodes strongly confined to the bulk of the layered material. Since the bulk modes originate from repulsion of GPPs, the total number of these modes is equal to the number of dielectric layers. The entire multilayer may also support surface modes at the interfaces with the claddings. In case of symmetric claddings they are similar to LRP and SRP of a metallic film.

The surface modes behave fundamentally different when $|\varepsilon_m|<\varepsilon_d$. Although a single interface does not support a SPP, a thin film of negative-permittivity medium supports two modes with identical field distribution ($TM_0$) but different "handedness" – the relation between the directions of electric and magnetic fields and wavevector of the mode[2, 13] (Fig. 1, right). One of these modes is a direct "right-handed" analog of a LRP. The other one is its negative-index "left-handed" counterpart. While the analogy is not complete, negative index medium is often called optical anti-matter.[14] The close-to-cut-off $TM_0$ mode doublet can therefore be considered as optical mode/anti-mode pair. Further analogy is seen in the formation of an *anti-symmetric* left-handed gap plasmon polariton in a dielectric gap between metallic claddings when $|\varepsilon_m|<\varepsilon_d$.[2, 15, 16] Similar to the case of $|\varepsilon_m|>\varepsilon_d$, absolute value of the modal index for the left-handed waves increases as the metal interfaces get closer. Naturally, such an increase of the absolute value of the modal index is accompanied by stronger mode confinement.

Figs. 2 and 3 further summarize the modal indices of metal-dielectric composites with 25nm layers and dielectric claddings, realized in our experiment. While optical losses of the bulk plasmon polariton modes in multilayers are larger than losses of SPPs these modes are of significant interest to nanophotonics due to extremely strong field confinement. Besides light guiding by sub-wavelength structures, nanoscale multilayers with appropriately patterned films are promising candidates for development of metamaterials with negative refractive index,[16-21] as well as for development of nano-sensors.

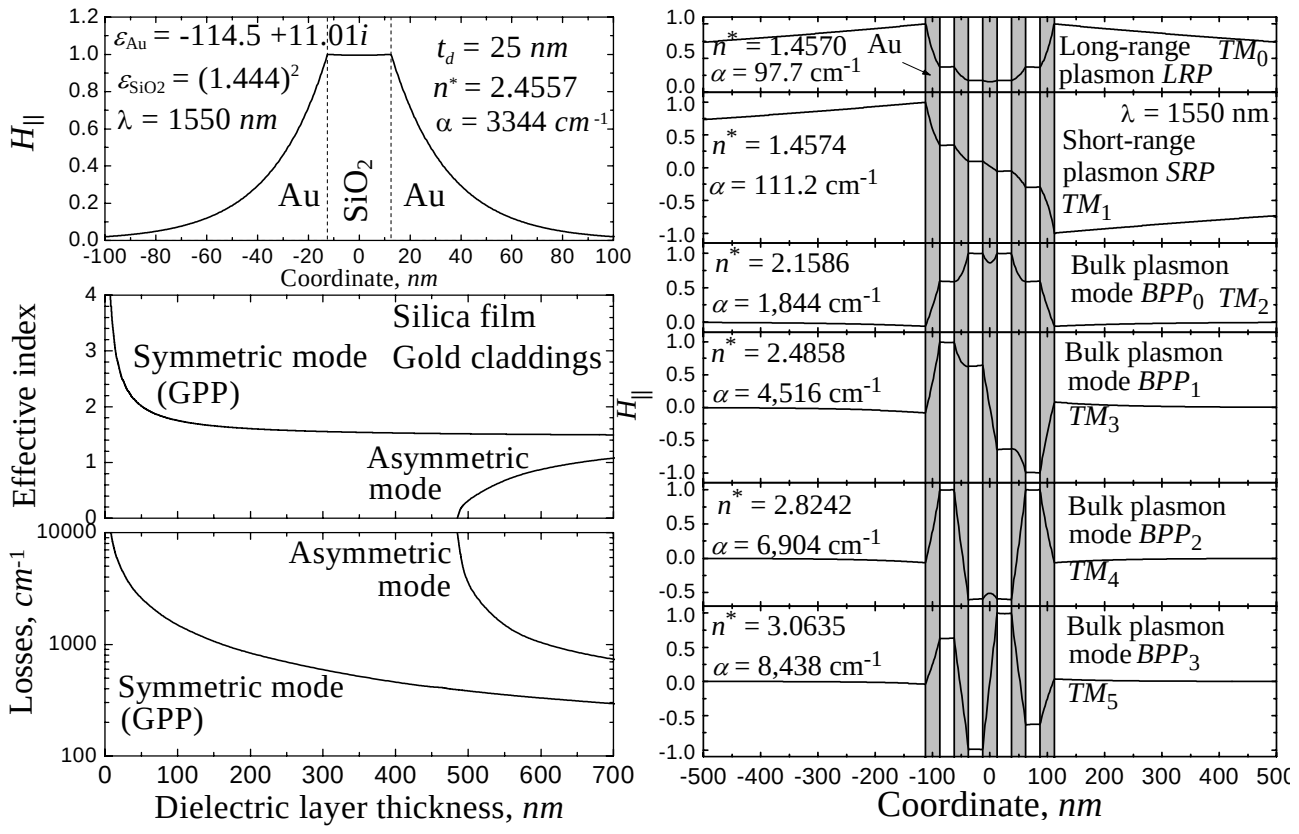

FIG. 2. Field profile (strength of magnetic field versus coordinate across the structure) for a gap plasmon polariton and its modal index and losses as a function of the dielectric layer thickness (left). Profiles of modes supported by a nanoscale multilayer of 4 silica layers and 5 gold layers between silica claddings (right).

The traditional classification scheme for transverse electromagnetic modes in multilayers relies on the number of nodes in the field distributions. Accordingly, the SPP supported by a single interface is a $TM_0$ mode. The LRP and SRP supported by a thin metal film are labeled as $TM_0$ and $TM_1$ modes. The gap plasmon polariton is a $TM_0$ mode. Such a mode labeling scheme is perfect for the dielectric waveguides but causes some confusion when applied to nanoscale metal-dielectric structures. It is a bit inconvenient that the same label $TM_0$ is designated to rather different electromagnetic excitations such as surface plasmon polaritons, long-range film plasmon polaritons, and the gap plasmon polaritons. Two modes of a metal strip with $|\varepsilon_m|<\varepsilon_d$ are both $TM_0$ waves. Further inconvenience is that essentially the same mode – the gap plasmon polariton – is labeled $TM_0$ in the structure with metallic claddings, and it becomes a $TM_2$ mode in a structure with dielectric claddings, two thin metallic layers, and the guiding dielectric layer. The $TM_n$ label indicates the number of nodes in the modal field distribution. This number, however, is not associated directly with the character of a particular mode guided by a nanoscale multilayer (smooth, oscillating, confined to the bulk of the multilayer, surface wave e.t.c.).

Once the highly confined bulk modes and the film-plasmon-polariton-type modes have distinct properties, it is reasonable to label them differently. In particular, in a structure with large number of layers, the bulk mode with relatively smooth profile is reasonable to call the fundamental bulk plasmon polariton mode $BPP_0$. In the traditional numbering scheme this is a $TM_2$ mode because it has two nodes close to the interfaces with the claddings. Accordingly, the bulk mode of order $n$ ($BPP_n$) would be labeled as $TM_{n+2}$ in the traditional classification. The labels such as *LRP* and *SRP* should be reserved for the modes with intensity maxima at the interfaces with the claddings. When claddings have different permittivities, the labels *LRP* and *SRP* become rather senseless. Instead, titles such as SPP bounded to particular interfaces will be more appropriate.

We stress that in a finite thickness nano-layered film the bulk modes, due to minor penetration into the claddings, are rather independent of the cladding indices. Therefore, all $BPP_n$ mode labeling is also independent of the claddings – a significant advantage over traditional scheme, which is strongly affected by claddings' indices.

The proposed scheme is illustrated in Fig. 2 (right) on the example of modes of a composite with $N_d$=4 dielectric and $N_m$=5 metallic 25nm thick layers. The structure supports the long-range plasmon polariton ($LRP=TM_0$), the short-range plasmon polariton ($SRP=TM_1$), and four bulk plasmon polariton modes ($BPP_0 \ldots BPP_3 = TM_2 \ldots TM_5$). Both LRP and SRP show large penetration into the claddings. In contrast, the bulk plasmon polariton modes are confined within the multilayer with minor fraction of optical power propagating in the claddings. The fundamental bulk mode $BPP_0$ has rather smooth field profile. For the highest order bulk mode, the modal field reveals fast oscillations so that it has opposite signs in neighboring dielectric layers. Note that the modal indices of the BPP modes can be several times higher than the refractive index of the dielectric in the multilayer. The origin of this surprising behavior is in the coupling-induced repulsion of the modal indices of individual gap plasmon polaritons discussed above.

The modal indices of guided modes in a multilayer with alternating 25nm thick layers of gold and silica are shown in Fig. 3. As predicted, the number of bulk modes is equal to the number of dielectric layers, and the maximal modal index increases with the number of layers increasing. For any given number of layers, modes of higher order have larger losses and larger modal indices. Assuming the number of layers is approaching infinity $N \to \infty$, we find the dispersion relation for the highest and the lowest order bulk plasmon polariton modes ($BPP_N$ and $BPP_0$) by setting periodical boundary conditions and requiring that the magnetic field strength has a node in the middle of every metal layer ($BPP_N$) or does not have such a node ($BPP_0$):

$$\tanh\left(\frac{\pi t_m}{\lambda}\sqrt{n^{*2}_{BPP_N}-\varepsilon_m}\right)\tanh\left(\frac{\pi t_d}{\lambda}\sqrt{n^{*2}_{BBP_N}-\varepsilon_d}\right)+\frac{\varepsilon_d}{\varepsilon_m}\sqrt{\frac{n^{*2}_{BPP_N}-\varepsilon_m}{n^{*2}_{BPP_N}-\varepsilon_d}}=0 \quad (2a)$$

$$\tanh\left(\frac{\pi t_d}{\lambda}\sqrt{n^{*2}_{BPP_0}-\varepsilon_d}\right)\Big/\tanh\left(\frac{\pi t_m}{\lambda}\sqrt{n^{*2}_{BPP_0}-\varepsilon_m}\right)+\frac{\varepsilon_d}{\varepsilon_m}\sqrt{\frac{n^{*2}_{BPP_0}-\varepsilon_m}{n^{*2}_{BPP_0}-\varepsilon_d}}=0 \quad (2b)$$

Eqs. (2a-b) do not contain the cladding indices or the overall composite thickness, further indicating "bulk" origin of these modes. In the limit of thin (nanoscale) layers, Eqs. (2a-b) yield following approximation for the modal indices

$$n^*_{BPP_N} \approx \sqrt{\varepsilon_d - \frac{\lambda^2}{\pi^2 t_d t_m}\frac{\varepsilon_d}{\varepsilon_m}} \qquad n^*_{BPP_0} \approx \sqrt{\frac{\varepsilon_d\varepsilon_m(t_d+t_m)}{t_d\varepsilon_m + t_m\varepsilon_d}}. \quad (3)$$

Note that the wavelength disappears from the expression for the modal index of the fundamental mode and the equation becomes equivalent to the predictions of the effective medium theory. The highest order mode has larger modal index than a single gap plasmon polariton provided that $t_{d,m} << \lambda/\pi\sqrt{|\varepsilon_m|}$. A particular case of the fundamental mode, the one with the smallest losses, was studied[22] back in 50's. SPPs supported by a structure consisting of few metal and dielectric thin films have been studied earlier.[23] Non-local corrections to the averaged permittivity in metal-dielectric nano-composites are reported in our recent paper.[24] Similar effect has been found in GHz systems with conducting wire-shaped inclusions.[25]

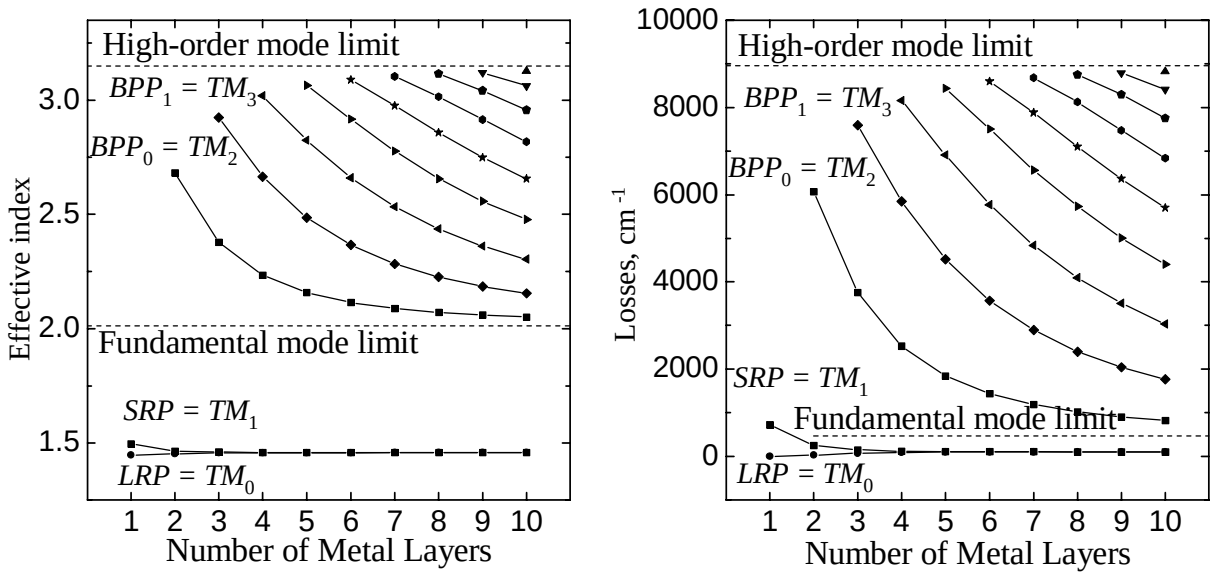

FIG. 3. Effective indices (left) and losses (right) for the modes in silica/gold multilayers. Dashed lines show the limits for the lowest order and highest order bulk modes in approximation of infinite number of layers (2).

For the gold-silica multilayers with $t_d$=$t_m$=25nm, Eq. (2) gives $\mathrm{Re}(n^*_{BPP_N})$= 3.1488 and $\alpha = (4\pi/\lambda)\mathrm{Im}(n^*_{BPP_N})$ =8,959cm$^{-1}$ – maximal possible values for the modal index and losses for the high-order bulk mode $BPP_N$. Corresponding values for the fundamental mode are $\mathrm{Re}(n^*_{BPP_0})$=2.0128 and $\alpha$= $(4\pi/\lambda)\mathrm{Im}(n^*_{BPP_0})$=462.8 cm$^{-1}$. These limits are shown in Fig. 3 by horizontal dashed lines.

The modes in a nanoscale metal-dielectric multilayer are relatively easy to predict and simulate numerically, but their experimental verification is challenging due to the deep sub-wavelength confinement and very high optical losses. In this paper we report, to the best of our knowledge, first experimental studies of guided modes in nanoscale metal-dielectric multilayers. To excite the high-index modes, we used the evanescent light coupling scheme. High-index material (silicon, $n_{Si}$=3.48) was used to match the wavevector of light in free space to the wavevector of a guided mode. To access wider range of modal indices, semi-cylinder geometry instead of more traditional prism was used. In a similar manner, a high-index semi-spherical solid immersion lens can be used, but for accurate angular measurement the semi-cylindrical geometry is preferable. Light from a fiber-coupled tunable (1490–1590nm) semiconductor laser (Photonetics Inc.) was collimated using a 10× objective. The laser was tuned to the wavelength of 1550nm, which was verified by an optical spectrum analyzer (Hewlett-Packard HP70951B). The spectral width of the laser radiation was below the resolution of the spectrometer (<0.1nm). Angular reflection spectrum was measured, and datum were plotted as a function of the product $n_{Si}\cdot\mathrm{Sin}(\theta)$, where $\theta$ is the incident angle. In this scale, the intensity minima directly indicate the modal indices of the guided modes excited through the evanescent coupling.

The multilayer structure was designed to consist of three pairs of silica(~25nm)/gold(~25nm). The layers were deposited directly on the flat facet of the semi-cylinder. The gold layers were deposited by electron beam evaporation, and silica layers – by plasma-enhanced chemical vapor deposition. With two dielectric gaps between three metal layers, the structure supports two BPP modes. The thickness of the layers was chosen to ensure that the effective indices of the bulk modes are in the comfortable for the measurement range ($n^*$<3.0). The first silica layer between silicon and gold is not crucial. Its role is to adjust the evanescent coupling strength for clear observation of guided modes. The experimental data are shown in Fig. 4 at the left. They verify guided modes with modal indices 2.31 and 2.88 recognized as $BPP_0$ and $BPP_1$. The structure was designed for measuring the modal indices of the bulk modes, while the SPP at the interface with the substrate is over-damped, and the SPP at the interface with air has vanishing small evanescent coupling with the incident beam. By fitting the experimental data with numerically simulated angular reflection, the best fit structure has been identified as Si/33nm-silica/24nm-gold/24nm-silica/26nm-gold/31nm-silica/24nm-gold (in average, 29nm-silica/25nm-gold), which is close to the deposition target numbers. Fig. 4 also shows the modes' profiles in the manufactured structure calculated using the best fit data for the layers' thicknesses.

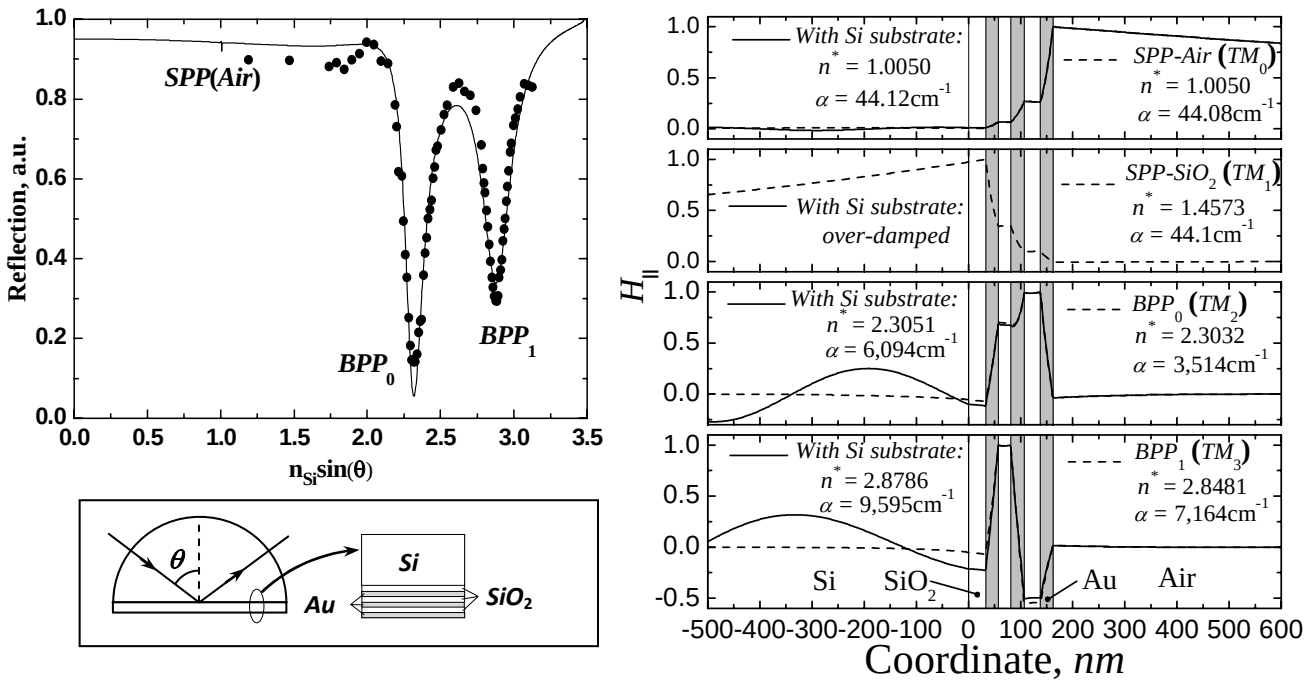

FIG. 4. Experimental reflection data for a silica/gold nanoscale multilayer (circles) and theoretical fit (solid line) indicating $BPP_0$ and $BPP_1$ modes (left). Simulated mode profiles in the experimental structure using the best fit thicknesses of the layers (right). Solid lines correspond to the structure with high-index silicon substrate. Dashed lines show mode profiles assuming the substrate is silica.

In conclusion, we analyzed *TM*-polarized modes supported by nanoscale metal-dielectric multilayers. We show that, in addition to the long-range and short-range plasmons supported by thin films, there are high-index guided modes strongly confined within the bulk of the multilayer. The dispersion relation is derived for the highest order and lowest order modes in approximation of infinite number of layers. New classification scheme is proposed for the modes supported in nanoscale metal-dielectric multilayers. High-index highly confined modes in a structure made of thee pairs of gold(~25nm)/silica(~29nm) layers are experimentally verified.

---

* Corresponding author. E-mail avrutsky@eng.wayne.edu